\documentclass{article}

\usepackage{arxiv}
\usepackage{amsthm} 
\usepackage{amssymb}
\usepackage{amsmath}
\usepackage{empheq}
\usepackage{amscd}
\usepackage{graphicx}
\usepackage{graphics}
\usepackage[noadjust]{cite}
\usepackage{amsthm}
\usepackage{caption}
\usepackage{multirow}
\usepackage{array}
\usepackage{rotating}
\usepackage[norelsize,ruled]{algorithm2e}
\usepackage{hyperref}

\usepackage{indentfirst}
\usepackage{tikz}
\usepackage{calc}
\usepackage{epsfig}
\usepackage[numbers]{natbib}
\usepackage[makeroom]{cancel}
\usepackage{multicol}
\usepackage{csquotes}
\usepackage{enumitem}

\usepackage{hyperref}       
\hypersetup{colorlinks,linkcolor={green},citecolor={green},urlcolor={black}}
\usepackage{optidef}

\begin{document}
\title{Fast Azimuthally Anisotropic 3D Radon Transform by Generalized Fourier Slice Theorem}

\author{
  \href{https://orcid.org/0000-0001-9588-0896}
 {\includegraphics[scale=0.06]{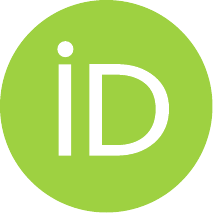}\hspace{1mm} AhmadReza Mokhtari} \\
  Institute of Geophysics, University of Tehran, Tehran, Iran 
  \And
\href{https://orcid.org/0000-0002-9879-2944}{\includegraphics[scale=0.06]{orcid.pdf}\hspace{1mm}Ali Gholami} \\
  Institute of Geophysics, Polish Academy of Sciences, Warsaw, Poland\\
  \texttt{agholami@igf.edu.pl} \\ 
  }



\renewcommand{\shorttitle}{Fast Azimuthally Anisotropic 3D Radon Transform ~~~~~~~~~~~~~~~~~~ Ali Gholami}

\maketitle
\begin{abstract}
Expensive computation of the conventional sparse Radon transform limits its use for effective transformation of
3D anisotropic seismic data cubes.
We introduce a fast algorithm for azimuthally anisotropic 3D Radon transform with sparsity constraints, allowing effective transformation of seismic volumes corresponding to arbitrary anisotropic inhomogeneous media.
In particular, a 3D data (CMP) cube of time and offset coordinates is transformed to a 3D cube of intercept time, slowness, and azimuth.
The recently proposed generalized Fourier slice theorem is employed for very fast calculation of the 3D inverse transformation and its adjoint, which are subsequently used for efficient implementation of the sparse transform via a forward-backward splitting algorithm.
The new anisotropic transform improves the temporal resolution of the resulting seismic data.
Furthermore, the Radon transform coefficients allows constructing azimuthally dependent NMO velocity curve at any horizontal plane, which can be inverted for the medium anisotropic parameters.
Numerical examples using synthetic data sets are presented showing the effectiveness of the proposed anisotropic method in improving seismic processing results compared with conventional isotropic counterpart.
\end{abstract}


\graphicspath{{"./Fig/"}}

\section{Introduction}

Nowadays, 3D seismic surveys are widely conducted to map the subsurface both in land and offshore. Even thought 3D surveys map the subsurface more accurate than 2D counterparts particularly in the complicated geological structures, but processing of such data is more complicated and demands more effort \citep{biondi20063d}. Aside from high amount of data, presence of (apparent) azimuthal anisotropy is one of the major problems that limits applicability of 2D processing tools such as velocity analysis, NMO correction and migration in 3D surveys \citep{tsvankin2010seismic}.

Two sources for observed azimuthal anisotropy in seismic data has been considered, namely, intrinsic  and apparent azimuthal anisotropy \citep{shragge2012elliptical}. Intrinsic azimuthal anisotropy accounts for the medium containing vertically aligned fractures, inhomogeneous stresses, aligned grains in rock matrix and in fact, represents the variation of velocity by direction of wave propagation in an (anisotropic) medium. Apparent azimuthal anisotropy is caused by dipping layers, lateral inhomogeneity of velocity in complex 3D structures. Observed azimuthal anisotropy could be caused by one or a combination of apparent and intrinsic azimuthal anisotropy.

Two approaches exist for velocity and anisotropy parameter estimation in the presence of azimuthal anisotropy. First approach is to decompose data into different azimuths and treat each azimuth independent of others. Even thought this approach could handle fully azimuthal anisotropy but it's applicability is limited in the case of small fold and/or when azimuthal coverage is not full. To the best of the authors knowledge, there is no an efficient algorithm to perform this approach efficiently. Second approach
is to develop the processing tools for a simple horizontal transverse isotropic (HTI) medium \citep{tsvankin2011seismology}. \citet{hu2015fast} proposed a two step fast algorithm based on butterfly algorithm for velocity analysis in a simple HTI medium.

In this paper, we address the problems of velocity analysis, multiple attenuation and high resolution (NMO-free) stacking of seismic data in the presence of azimuthal anisotropy by developing hyperbolic Radon transform (HRT) for azimuthal anisotropic media. We develop a fast Azimuthal Anisotropic HRT (AAHRT) operator based on recently proposed generalized Fourier slice theorem (GFST) whose computational cost is $O(N\log N)$ where $N$ is the model/data size. Literature review indicates that due to high computational burden of AAHRT, there is no paper published to solve such problem via inverse theory. Taking the advantages of our proposed fast algorithm for computation of the canonical forward/adjoint AAHRT operators, we decompose a 3D CMP gather via sparse regularization to achieve high resolution coefficients in the transform domain.
Numerical tests on synthetic and real data confirm the high performance of the proposed algorithms for velocity analysis and reconstruction of seismic data in the presence of azimuthal anisotropy.

\begin{figure}
\centering
\includegraphics[width=14cm,height=5cm,trim={50cm 0 0 0},clip]{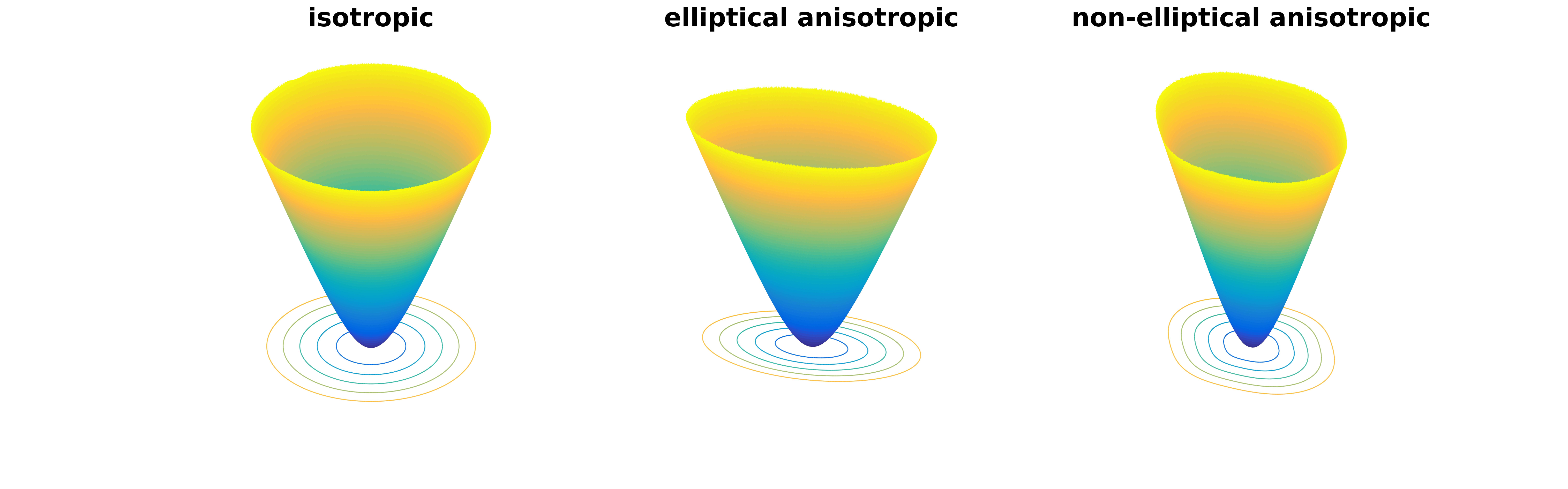}
\caption{Isotropic versus azimuthal anisotropic 3D hyperbolic Radon basis functions.}
\label{Fig_AAGFST}
\end{figure}
\section{Azimuthally anisotropic hyperbolic Radon transform (AAHRT)}
Let $D(x,y,t)$ indicate a 3D CMP data with temporal coordinate $t$ and spatial coordinates $x$ and $y$, then the generalized Radon transform allows the representation of $D(x,y,t)$ as a superposition of the generalized Radon basis functions  as \citep{gholami2017three}
\begin{eqnarray}
D(t,x,y) =\int_{\boldsymbol{\Omega}} M(\boldsymbol{\alpha}) \Psi_{\boldsymbol{\alpha}}(t,x,y)  ~d\boldsymbol{\alpha},
\label{main}
\end{eqnarray}
where $\boldsymbol{\alpha}\in \boldsymbol{\Omega}$ is a multiindex parameter, $\boldsymbol{\Omega}$ is the domain of the
parameter space, $M(\boldsymbol{\alpha})$ are the coefficients of the forward transform, and $\Psi_{\boldsymbol{\alpha}}(t,x,y)$ are transform basis functions. Although different forms of basis functions (such as linear and parabolic) can be considered, the focus of this paper is on hyperbolic basis functions  \citep[][ for more details]{gholami2017three,gholami2017deconvolutive}.  
Hyperbolic basis functions can be defined by the following general equation and are effective for decomposing data corresponding to a typical spread length equal to the reflector depth:
\begin{eqnarray}
\Psi_{\boldsymbol{\alpha}}(t,x,y)=\delta(t-\sqrt{\tau^2+q(\boldsymbol{\alpha})(x^2+y^2)}) .
\label{RT_basis}
\end{eqnarray}
where $\tau$ is zero-offset reflection time (intercept time) and $q(\boldsymbol{\alpha})$ is curvature parameter (squared slowness).
For an isotropic earth model, $q$ is constant, i.e. the wave velocity do not varies with horizontal direction of wave vector.
In this case, the basis functions are characterized by two parameters $\tau$ and $q$ and hence (isotropic) Radon transform maps the 3D data $\Psi_{\boldsymbol{\alpha}}(t,x,y)$ to a 2D  Radon space $M(\tau,q)$. Recently, efficient algorithms have been proposed for fast computation of such isotropic Radon transforms \citep{hu2013fast, nikitin2017fast, gholami2019fast,mokhtari2019fast}.

In reality, however, due to some factors such as presence of vertically aligned fractures, inhomogeneous stresses, lateral velocity variations, dipping layers or a combination of them, the value of parameter $q$ in eq. \eqref{RT_basis} may vary by azimuth of seismic line.
The velocity of a pure-mode P-wave propagating in a 3D medium that exhibits weak azimuthal anisotropy varies with azimuth $\theta=\arctan(y/x)$ as \citep{backus1965possible}
\begin{eqnarray}
 V(\theta) = \alpha_1 +\alpha_2 \cos(2\theta)+ \alpha_3 \sin(2\theta) + \alpha_4 \cos(4\theta)+ \alpha_5 \sin(4\theta)
\end{eqnarray}
where the five constants $\alpha_1,...,\alpha_5$ are a function of the 21 anisotropic elastic parameters of the material through which the waves are propagating.
Therefore, since $V\equiv q^{-2}$, the Radon space of a fully azimuthally anisotropic hyperbolic Radon transform (AAHRT) is 6D ($\tau$ + $\alpha_1,...,\alpha_5$).
This makes computation of the 6D transform described by eq. \eqref{main}, if not impossible, extremely expensive.

\subsection{Radon Transform with Elliptical Anisotropy}
In practical applications, a more simple form of azimuthal anisotropy that is of broad applicability is the so-called elliptical anisotropy (which is equivalent to transverse isotropy with a horizontal axis of symmetry). Figure \ref{Fig_AAGFST} compares isotropic and anisotropic (elliptical and non-elliptical) hyperbolic basis functions. 
In the case of elliptical anisotropy, azimuth-dependent curvature parameter is described by
\begin{eqnarray} \label{ellip1}
q(\theta) = q_x \cos(\theta)^2 + q_y \sin(\theta)^2
\end{eqnarray}
where $q_x$ and $q_y$ are the orthogonal components of curvature in $x$ and $y$ direction.
This simplifies the basis functions in eq. \eqref{RT_basis} to
\begin{eqnarray}
\Psi_{\tau,q_x,q_y}(t,x,y)=\delta(t-\sqrt{\tau^2+q_x x^2+q_y y^2}) .
\label{RT_basis}
\end{eqnarray}
with the corresponding Radon transform
\begin{eqnarray}
D(t,x,y) =\iiint M(\tau,q_x,q_y)\delta(t-\sqrt{\tau^2+q_x x^2+q_y y^2})~d\tau~dq_x~dq_y.
\label{main_ellip}
\end{eqnarray}
However, computation of this triple integral transform is still expensive.

\subsection{Radon Transform with General Anisotropy}
Generally, the elliptic assumption about anisotropy may be inadequacy for describing some models in nature \citep{thomsen1986weak}.
Therefore, in this paper, we consider a general model for azimuthally anisotropic Radon transform with the following basis functions
\begin{eqnarray} \label{basis}
\Psi_{\tau,q,\theta}(t,x,y)=\delta(t-\sqrt{\tau^2+qx^2+qy^2})~ \delta(\theta-\arctan(y/x)),
\end{eqnarray}
and, accordingly, we define the following general AAHRT model 
\begin{eqnarray}
D(t,x,y) =\iiint M(\tau,q,\theta)\delta(t-\sqrt{\tau^2+qx^2+qy^2})~ \delta(\theta-\arctan(\frac{y}{x})) ~d\tau~dq~d\theta .
\label{Inverse}
\end{eqnarray}
Equation \eqref{Inverse} defines a linear mapping between the 3D CMP data $D(x,y,t)$ of spacial coordinates $x$ and $y$ and temporal coordinate $t$ and the 3D AAHRT coefficients $M(\tau,q,\theta)$ of intercept time $\tau$, curvature $q$ and azimuth $\theta$. 

\section{Fast Calculations of Forward/Inverse Operators}
In this section, we develop very fast algorithms for implementation of the forward and inverse 3D Radon transforms with arbitrary anisotropy.

\subsection{Fast Inverse Operator}
We provide a very fast algorithm for computation of eq. \eqref{Inverse} via the generalized Fourier slice theorem (GFST) \citep{gholami2017time}.
The GFST is a method for fast implementations of time-invariant Radon transforms. For 2D data, GFST reduces the computational cost of time-invariant Radon transforms from $O(N^3)$ (for direct calculations) to $O(N\log N)$, where $N$ is number of data/model  samples. 
In order to apply the GFST to our problem in this paper, the first step is to make the basis functions time-invariant. This can be done by using the old method of time stretching \citep{yilmaz1989velocity,nikitin2017fast, gholami2019fast,mokhtari2019fast}. 
Accordingly, a change of variable $t' \leftarrow t^2$ and  $\tau' \leftarrow \tau^2$ in eq. \eqref{Inverse} leads to
\begin{eqnarray}
D(t',x,y) =\iiint M(\tau',q,\theta)\delta(t'-\tau'-qx^2-qy^2)~ \delta(\theta-\arctan(y/x)) ~d\tau'~dq~d\theta.
\label{Inverse2}
\end{eqnarray}
Equation \eqref{Inverse2} is simply a 3D  azimuthally anisotropic parabolic Radon transform (AAPRT) and thus suitable to be implemented by the GFST.
Applying the 1D Fourier transform to both sides of eq. \eqref{Inverse2}, with respect to $\tau'$, yields
\begin{eqnarray}
D(f',x,y) &=& \iiint M(\tau',q,\theta)~ \delta(\theta-\arctan(y/x)) \left\{\int \delta(t'-\tau'-qx^2-qy^2) e^{-i2\pi f't'}  ~dt'\right\}~d\tau'~dq~d\theta,\\
&=& \iint M(\tau',q,\theta)~ \delta(\theta-\arctan(y/x)) e^{-i2\pi f'(\tau'+qx^2+qy^2)}  ~d\tau'~dq~d\theta,\\
&=& \int \delta(\theta-\arctan(y/x)) \left\{ \int \left\{ \int M(\tau',q,\theta) e^{-i2\pi f'\tau'} ~d\tau' \right\} e^{-i2\pi f'(qx^2+qy^2)} ~dq \right\}  ~    ~d\theta,\\
&=& \int \delta(\theta-\arctan(y/x)) \left\{ \int  \widehat{M}(f',q,\theta) e^{-i2\pi f'q(x^2+y^2)} ~dq \right\}  ~    ~d\theta,\\
&=& \int \delta(\theta-\arctan(y/x)) \widehat{M}(f',k_q=f'x^2+qy^2),\theta)   ~    ~d\theta,\\
&=&  \widehat{M}(f',k_q=f'x^2+qy^2,\theta=\arctan(y/x)), \label{Inv_op} 
\end{eqnarray}
where  $\widehat{M}(f',k_q,\theta)$ is 2D Fourier transform of ${M}(\tau',q,\theta)$ and $ \widehat{D}(f',x,y)$ is 1D Fourier transform of $ {D}(t',x,y)$.
According to eq. \eqref{Inv_op} the coefficients $ \widehat{D}(f',x,y)$ are related to $\widehat{M}(f',k_q,\theta)$  by the equation
\begin{eqnarray} \label{inv_GFST}
\begin{cases}
k_q = f'x^2+f'y^2,\\
\theta=\arctan(\frac{y}{x}).
\end{cases}
\end{eqnarray}
A schematic representation of azimuthal anisotropic GFST in eq. \eqref{inv_GFST} is shown in figure \ref{Fig_AAGFST}. The left cube is CMP data in $(f',x,y)$ domain and the right cube is the corresponding Radon cube in $(f',k_p,\theta)$ domain. The white curves in each cube are define by eq. \eqref{inv_GFST}. The data in the specified azimuth $\theta$ are mapped to the constant $\theta$ section in the Radon domain. \\
\textbf{Algorithm 1.} The proposed workflow for calculating the inverse transform includes the following five steps:
\begin{itemize}
\item[Step 1.] Stretch the time axis of the Radon cube via an interpolation method, ${M}(\tau,q,\theta)\rightarrow {M}(\tau',q,\theta)\equiv {M}(\tau^2,q,\theta)$.
\item[Step 2.] Apply the 2D Fourier transform along $\tau'$ and $q$, ${M}(\tau',q,\theta)\rightarrow \widehat{M}(f',k_q,\theta)$.
\item[Step 3.] Interpolate $\widehat{M}(f',k_q,\theta)$ to estimate the $f'-x-y$ coefficients $\widehat{D}(f',x,y)$  according to eq. \eqref{inv_GFST}. 
\item[Step 4.] Apply the 1D inverse Fourier transform along $f'$, $\widehat{D}(f',x,y)\rightarrow {D}(t',x,y)$.
\item[Step 5.] Undo the time stretching operation, $D(t',x,y)\equiv D(t^2,x,y)\rightarrow D(t,x,y)$.
\end{itemize} 
\subsection{Fast Adjoint Operator (Fast 3D Azimuthally Anisotropic Velocity Stack)}
The adjoint operator of AAHRT in eq. \eqref{Inverse} reads as
\begin{eqnarray}
 M(\tau,q,\theta)=\iiint D(t,x,y)\delta(t-\sqrt{\tau^2+qx^2+qy^2})~ \delta(\theta-\arctan(y/x))  ~dt~dx~dy,
\label{Adjoint}
\end{eqnarray}
which is in fact the zero-lag cross-correlation of the CMP data and the hyperbolic basis functions, simillar to the well known 
velocity stack \citep{taner1969velocity}. Velocity stack is basically the adjoint solution of Radon transform. 
To the best of the authors knowledge there is no fast algorithm for computing 3D velocity stack with a general anisotropy as defined in eq. \eqref{Adjoint}. However, based on the fast butterfly method of \citet{hu2013fast}, \cite{hu2015fast} proposed a fast algorithm for velocity scan with elliptical anisotropy.

A change of variable $t' \leftarrow t^2$ and  $\tau' \leftarrow \tau^2$ in eq. \eqref{Adjoint} leads to
\begin{eqnarray}
M(\tau',q,\theta) =\iiint D(t',x,y)\delta ( t'-\tau'-qx^2-qy^2) ~ \delta(\theta-\arctan(y/x)) ~dt' ~dx~dy.
\label{Adjoints}
\end{eqnarray}
Applying the 2D Fourier transform to both sides of eq. \eqref{Adjoints}, with respect to $\tau'$ and $q$, yields
\begin{eqnarray}
\widehat{M}(f',k_q,\theta) &=& \iiiint D(t',x,y) \left\{\int \delta(t'-\tau'-qx^2-qy^2) e^{-i2\pi f'\tau'} ~d\tau'\right\}  e^{-i2\pi k_qq} \delta(\theta-\arctan(y/x)) ~ dq  ~dt'~dx~dy, \\
&=& \iiiint D(t',x,y) e^{-i2\pi f'(t'-qx^2-qy^2)} e^{-i2\pi k_qq}\delta(\theta-\arctan(y/x))   ~dq~ dt' ~dx~dy, \\
&=& \iint  \left\{\int D(t',x,y) e^{-i2\pi f't'} ~ dt'\right\}
\left\{\int e^{-i2\pi (k_q - f' x^2- f'y^2)q }  ~dq \right\} \delta(\theta-\arctan(y/x)) ~dx~dy, \\
&=&\iint  \widehat{D}(f',x,y) \delta(k_q-f'x^2-f'y^2)  \delta(\theta-\arctan(y/x))~  dx ~dy \\
&=& \widehat{D}(f',x = \sqrt{\frac{k_q}{f'}} \cos(\theta),y =\sqrt{\frac{k_q}{f'}} \sin(\theta)) \label{Fwd_op}
\end{eqnarray}
where  $\widehat{M}(f',k_q,\theta)$ is 2D Fourier transform of ${M}(t',q,\theta)$ and $ \widehat{D}(f',x,y)$ is 1D Fourier transform of $ {D}(t',x,y)$.
According to eq. \eqref{Fwd_op} the coefficients $\widehat{M}(f',k_q,\theta)$ are related to $ \widehat{D}(x,y,f')$ by the relation
\begin{eqnarray} \label{frwrd}
\begin{cases}
x = \sqrt{\frac{k_q}{f'}} \cos(\theta),\\
y =\sqrt{\frac{k_q}{f'}} \sin(\theta) .
\end{cases}
\end{eqnarray}
\textbf{Algorithm 2.} The proposed workflow for anisotropic velocity stack is performed in the following five steps:
\begin{itemize}
\item[Step 1.] Stretch the time axis of the CMP data via an interpolation method, ${D}(t,x,y)\rightarrow {D}(t',x,y)\equiv {D}(t^2,x,y)$.
\item[Step 2.] Apply the Fourier transform along the stretched axis, ${D}(t',x,y)\rightarrow \widehat{D}(f',x,y)$.
\item[Step 3.] Interpolate $\widehat{D}(f',x,y)$ to estimate the coefficients $\widehat{M}(f',k_q,\theta)$ according to eq. \eqref{frwrd}. 
\item[Step 4.] Apply the 2D inverse Fourier transform along $f'$ and $k_q$, $\widehat{M}(f',k_q,\theta)\rightarrow {M}(\tau',q,\theta)$.
\item[Step 5.] Undo the time stretch operation, ${M}(\tau',q,\theta)\equiv {M}(\tau^2,q,\theta)\rightarrow {M}(\tau,q,\theta)$.
\end{itemize} 

\subsection{Automatic Anisotropic Velocity Peaking}
The variations of velocity with azimuth can reveal useful information about subsurface and can help to better interpret seismic sections. Here we show how to automatically extract the stacking velocity as a function of time and azimuth.
Let $M$ be Radon cube generated by anisotropic velocity stack which contains only primary energy then the stacking velocity as a function of time and azimuth is estimated as \citep{gholami2017deconvolutive}
\begin{eqnarray}
\bold{v}^{-2}(\tau,\theta) = \frac{\int q M(\tau,q,\theta) dq}{\int M(\tau,q,\theta) dq}.
\label{vel_est}
\end{eqnarray}

The forward  and inverse Radon transforms defined in Eqs \eqref{Inv_op} and \eqref{Fwd_op} are formulated in the continuous domain. In the next subsection, discuss how to implement them in the discrete domain to process discrete data.

\begin{figure}
\centering
\includegraphics[width=14cm,height=6cm]{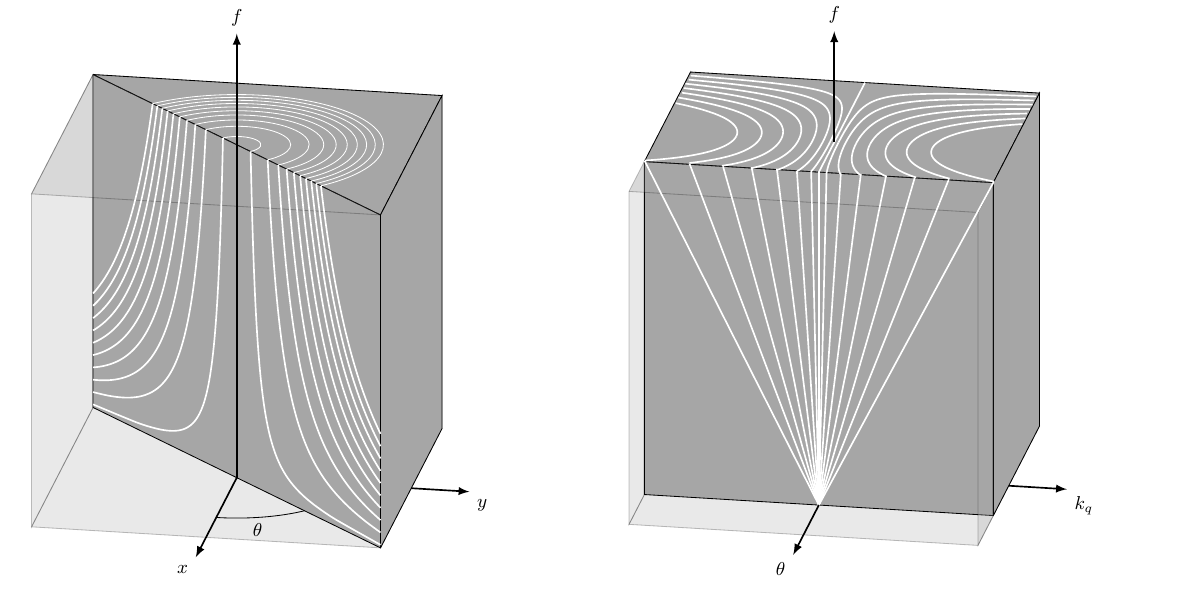}
\caption{Schematic representation of azimuthal anisotropic GFST (Eqs \eqref{inv_GFST} and \eqref{frwrd}).}
\label{Fig_AAGFST}
\end{figure}

%
%

\subsection{Discrete Implementation}
In order to apply the proposed method on discrete data, we have to rewrite AAHRT formulation in the discrete domain.
In this section, we show how to implement each step of Algorithms 1 and 2 in discrete domain. 

Step 1 in both Algorithms 1 and 2 is due to time stretching operation. This step can be easily performed by an interpolation method such as the spline interpolation via a linear operator $\bold{S}$. Step 5 of these algorithms reverses the time stretching operation via the inverse operator $\bold{S}^{-1}$. The interpolation is a local method and thus both $\bold{S}$ and $\bold{S}^{-1}$ can be performed with a computational complexity proportional to the number of grid points.  

Steps 2 and 4 in both Algorithms 1 and 2 requires discrete Fourier transformations (1D and 2D) on uniform grids this both steps can be carried out by fast Fourier transform (FFT) algorithms.   

Step 3 of both algorithms requires an interpolation to map the coefficient from Cartesian grid points to non-Cartesian grid points defined  according to Eqs \eqref{inv_GFST} or \eqref{frwrd}. We use linear interpolation for its efficiency and sufficient accuracy.
Consider the inverse transformation which requires computation of $\widehat{{D}}{(f',x,y)}$ from $\widehat{{M}}{(f',k,\theta)}$ using the relation \eqref{inv_GFST}. Since $f'$ appears in both variables $\widehat{{D}}$ and $\widehat{{M}}$ there is no need to interpolate in $f'$ direction. 
Assume that $k$ and $\theta$ are discretized uniformly, that is, $k_i=i\Delta k$ and $\theta_j=j\Delta \theta$ for $i,j=0,1,...$.
Then the coefficients $\widehat{{D}}{(f',x,y)}$ for a three-tuple $(f,x,y)$ are linearly related to the coefficients in $(f',k_i,\theta_j)$ domain as
\begin{eqnarray}
\begin{array}{c}
\widehat{D}(f',x,y) = 
A_1\widehat{M}(f',k_i,\theta_j) + 
A_2\widehat{M}(f',k_i,\theta_{j+1}) + 
A_3\widehat{M}(f',k_{i+1},\theta_j) +
A_4\widehat{M}(f',k_{i+1},\theta_{j+1})
\end{array}
\label{thli}
\end{eqnarray}
where the interpolation coefficients are
\begin{eqnarray} 
\begin{cases}
A_1=\frac{1}{\Delta k \Delta\theta}   \left(k_{i+1} - f'x^2-f'y^2\right) \left(\theta_{j+1} - \arctan(y/x) \right),\\
A_2=\frac{-1}{\Delta k \Delta\theta}  \left(k_{i+1} - f'x^2-f'y^2 \right) \left(\theta_j- \arctan(y/x) \right), \\
A_3=\frac{-1}{\Delta k \Delta\theta} \left(k_i -f'x^2-f'y^2\right) \left(\theta_{j+1} - \arctan(y/x) \right), \\
A_4=\frac{1}{\Delta k \Delta\theta}   \left(k_i  - f'x^2-f'y^2\right) \left(\theta_j-\arctan(y/x)\right),
\end{cases}
\end{eqnarray}
and the integers $i$ and $j$ are determined such that the following inequalities are satisfied 
\begin{eqnarray} 
i\Delta k \leq  f'x^2+f'y^2 \leq (i+1)\Delta k,\hspace{1cm }
j\Delta \theta\leq \arctan(y/x) \leq (j+1)\Delta \theta.
\end{eqnarray}

Similarly, if $x$ and $y$ are discretized uniformly, that is, $x_m=m\Delta x$ and $y_n=n\Delta y$ for $m,n=0,1,...$.
Then the coefficients $\widehat{{M}}{(f',k,\theta)}$  for a three-tuple $(f',k,\theta)$ are linearly related to the coefficients in $(f',x_m,y_n)$ domain as
\begin{eqnarray}
\begin{array}{c}
\widehat{M}(f',k,\theta) = 
B_1\widehat{D}(f',x_m,y_n) + 
B_2\widehat{D}(f',x_m,y_{n+1}) + 
B_3\widehat{D}(f',x_{m+1},y_n) +
B_4\widehat{D}(f',x_{m+1},y_{n+1})
\end{array}
\label{thli}
\end{eqnarray}
where the interpolation coefficients are
\begin{eqnarray} 
\begin{cases}
B_1=\frac{1}{\Delta x \Delta y}   \left(x_{m+1} - \sqrt{\frac{k}{f'}}\cos(\theta)\right) \left(y_{n+1} - \sqrt{\frac{k}{f'}}\sin(\theta) \right),\\
B_2=\frac{-1}{\Delta x \Delta y}  \left(x_{m+1} - \sqrt{\frac{k}{f'}}\cos(\theta) \right) \left(y_n- \sqrt{\frac{k}{f'}}\sin(\theta) \right), \\
B_3=\frac{-1}{\Delta x \Delta y} \left(x_m -      \sqrt{\frac{k}{f'}}\cos(\theta) \right) \left(y_{n+1} - \sqrt{\frac{k}{f'}}\sin(\theta) \right), \\
B_4=\frac{1}{\Delta x \Delta y}   \left(x_m  -    \sqrt{\frac{k}{f'}}\cos(\theta) \right) \left(y_n-\sqrt{\frac{k}{f'}}\sin(\theta)\right),
\end{cases}
\end{eqnarray}
and the integers $m$ and $n$ are determined such that the following inequalities are satisfied 
\begin{eqnarray} 
m\Delta x\leq  \sqrt{\frac{k}{f'}}\cos(\theta) \leq (m+1)\Delta x,\hspace{1cm }
n\Delta y\leq \sqrt{\frac{k}{f'}}\sin(\theta) \leq (n+1)\Delta y.
\end{eqnarray}

\subsection{The Range of Curvature Parameter}
The value of $q$ is bounded in the interval $[-q_{max}~~q_{max}]$ as we use FFT algorithm. In order to avoid aliasing along the $q$ axis, $q$ and hence $k_q$ should be sampled properly. Assuming $N_q$ equally spaced $q$ values, then sampling interval for $k$ is \citep{gholami2017time},
 \begin{eqnarray}
 \Delta k=\frac{f'_{max} (x_{max}^2+y^2_{max})}{N_q},
\label{q1}
\end{eqnarray}
where $f'_{max}$ is the maximum frequency to be inverted and and $x_{max}/y_{max}$ is the maximum offset is $x/y$ direction. According to the Nyquist sampling theorem we have that
 \begin{eqnarray}
 q_{max} = \frac{1}{2\Delta k}.
\label{q2}
\end{eqnarray}
Therefore, in order to avoid aliasing we need to satisfy
\begin{eqnarray}
 N_q \geq 2 q_{max} f'_{max} (x_{max}^2+y^2_{max}).
\label{q3}
\end{eqnarray}

In practice, $q_{max}$ is determined according to the steepest event in the section and using eq. \eqref{q3}. Constraining the $q$ values to a predefined range and reducing the number of curvature samples $N_q$ leads to a decrease in computational burden of the algorithm. For this purpose, we apply a partial NMO to the time-stretched data before Radon transformation.

\section{Sparsity-promoting AAHRT}
Discretization of the AAHRT formula and restricting the basis functions to a finite-dimensional vector space (to make them suitable for decomposition of real seismic data with finite spatial aperture) limits the resolution of the transform. 
 The generalized inverse operator maps the data from the Radon space to the physical data space.  The forward operator is however nonunique that means that there exists different Radon coefficients $M(\tau,q,\theta)$ that explain the data equally well \citep{beylkin1987discrete}. 
Thus, in practice, the forward Radon transform is usually cast as an inverse problem, allowing  additional constraints (such as sparsity) to be imposed in the Radon domain in order to render the forward solution unique. 
Following the steps given in Algorithm 1, the forward transform can be cast as the solution of an inverse problem of the form
\begin{eqnarray}
\bold{d}= \bold{S^{-1}}\bold{F}_1^{-1} \bold{P} \bold{F}_2\bold{S} \bold{m}
\label{axb2}
\end{eqnarray}
where $\bold{d}$ is the data vector, $\bold{m}$ is the model vector (AAHRT coefficients), $\bold{S^{-1}}$ are respectively the the spline interpolation operator and its inverse. $\bold{F}_2$ is the 2D discrete Fourier transform operator, $\bold{F}_1^{-1}$ is the 1D inverse discrete Fourier transform operator, and $\bold{P}$ is an interpolation operator defined  according to eq. \eqref{inv_GFST}. 
%
We define the solution $\bold{m}$ of eq. \eqref{axb2} as 
\begin{eqnarray}
 \arg\min_{\bold{m}}~~ \frac{1}{2}\|\bold{d}-\bold{R^{-1}}  {\bold{m}}\|_2^2 + \lambda \|\bold{m}\|_1,
\label{cost1}
\end{eqnarray}
where $\bold{R^{-1}}\equiv \bold{S^{-1}}\bold{F}_1^{-1} \bold{P} \bold{F}_2\bold{S}$.
The first term in eq. $\eqref{cost1}$ is the data fidelity. It ensures that the recovered CMP
is close to the observed one in the $\ell_2$ norm sense, by which we assume that the error in the data are Gaussian distributed random. The second term is the regularization, which encodes the prior knowledge about the model space. Here we use the  sparsity-promoting $\ell_1$ norm (sum of the absolute value coefficients) as the prior. This regularizer enforces the sparseness of the coefficients, i.e. fitting the data by the least number of basis functions.
In eq. $\eqref{cost1}$, $\lambda>0$ is the trade-off parameter which balances between data fitting and sparseness of the coefficients.

 We employ the proximal forward-backward splitting algorithm \citep{gholami2011general} to carry out the optimization described by eq. \eqref{cost1}. This algorithm finds the solution by a two steps iteration  
\begin{eqnarray} \label{ISTA}
\bold{m}^{k+1} = \mathcal{S}_{\lambda}(\bold{m}^{k} + \gamma \bold{R}^{-T}(\bold{d}-\bold{R}^{-1}\bold{m}^k)),
\end{eqnarray}
where $\mathcal{S}_{\lambda}$ is the well-known soft shrinkage operator, defined as
\begin{eqnarray}
\mathcal{S}_{\lambda}(x) = x \max(1 - \frac{\gamma\lambda}{|x|},0).
\label{soft}
\end{eqnarray}
In eq. \eqref{ISTA}, $\bold{R}^{-T}$ is the adjoint of $\bold{R}^{-1}$ and $\gamma>0$ is the step length.

\subsection{Deconvolutive AAHRT}
A main issue with the focusing power of the AAHRT discussed above is due to the shape of the basis functions $\Psi_{\tau,q,\theta}(t,x,y)$ (eq. \eqref{basis}) in the time direction. As seen the basis functions are Dirac delta functions in time, whereas the seismic events are band-limited due to the effects of the seismic wavelet. In order to increase the time resolution of the transform, \citet{gholami2017deconvolutive} proposed the so-called deconvolutive Radon transform by replacing the Delta function $\delta(t)$ in the definition of the basis functions with a function $w(t)$ representing the band-limited seismic wavelet.  
The basis functions for AAHRT read as
\begin{eqnarray}
\Psi_{\tau,q,\theta}(t,x,y)=w(t-\sqrt{\tau^2+qx^2+qy^2})~ \delta(\theta-\arctan(y/x)),
\end{eqnarray}
which incorporates into eq. \eqref{cost1} by replacing $\bold{R^{-1}}$ with  
\begin{eqnarray}
\bold{R^{-1}_w} = \bold{W} \bold{R^{-1}}
\label{DecRT}
\end{eqnarray}
where $\bold{W}$ is a convolution operator which performs linear convolution of the wavelet and the CMP in time direction.

\section{Numerical tests}
In this section, we test the performance of the proposed AAHRT algorithms using both synthetic and real CMP data sets. Using numerical tests, we show the computational cost of the forward/inverse transform in comparison with a direct calculation and resolution enhancement gained by the sparsity-promoting AAHRT and its deconvolutive version. In particular, we show that azimuthally anisotropic 3D velocity scan of very large CMP data is possible by the proposed algorithm in a reasonable time. 
Very fast computation of the forward/inverse transforms allows performing the iterative sparse solver given by eq. \eqref{ISTA} even on large 3D CMP data, which is suitable for anisotropic 3D data interpolation.

\subsection{Azimuthally Anisotropic Velocity Stack}
We use a synthetic CMP data shown in Fig. \ref{Fig_syn_3d} in which $dt=0.004$ s and $dx=dy=40$ m. The CMP is composed of a set of hyperbolic events where the associated anisotropic velocities are sown in Fig. \ref{Fig_vel_compare}a. 
We computed the anisotropic velocity scan in \eqref{Adjoints} by directly computing the associated Riemann sum and also by the proposed algorithm given in Algorithm 2.  
The results are shown in Figs \ref{Fig_RTs}a-b and their difference in Fig. \ref{Res_direct_proposed}. Little different can be seen between the two velocity cube while the computation time for performing the direct summation method (Fig. \ref{Fig_RTs}a) was significantly more than that of the proposed fast method (Fig. \ref{Fig_RTs}b). 
Using data sets of varying sizes, we quantitatively compared the calculation times of these approaches in Table 1. As expected, the proposed method performs significantly faster than direct summation. 
 
\subsection{Azimuthally Anisotropic Sparse Velocity Stack}
\begin{figure}
\centering
\includegraphics[width=11cm,height=8cm]{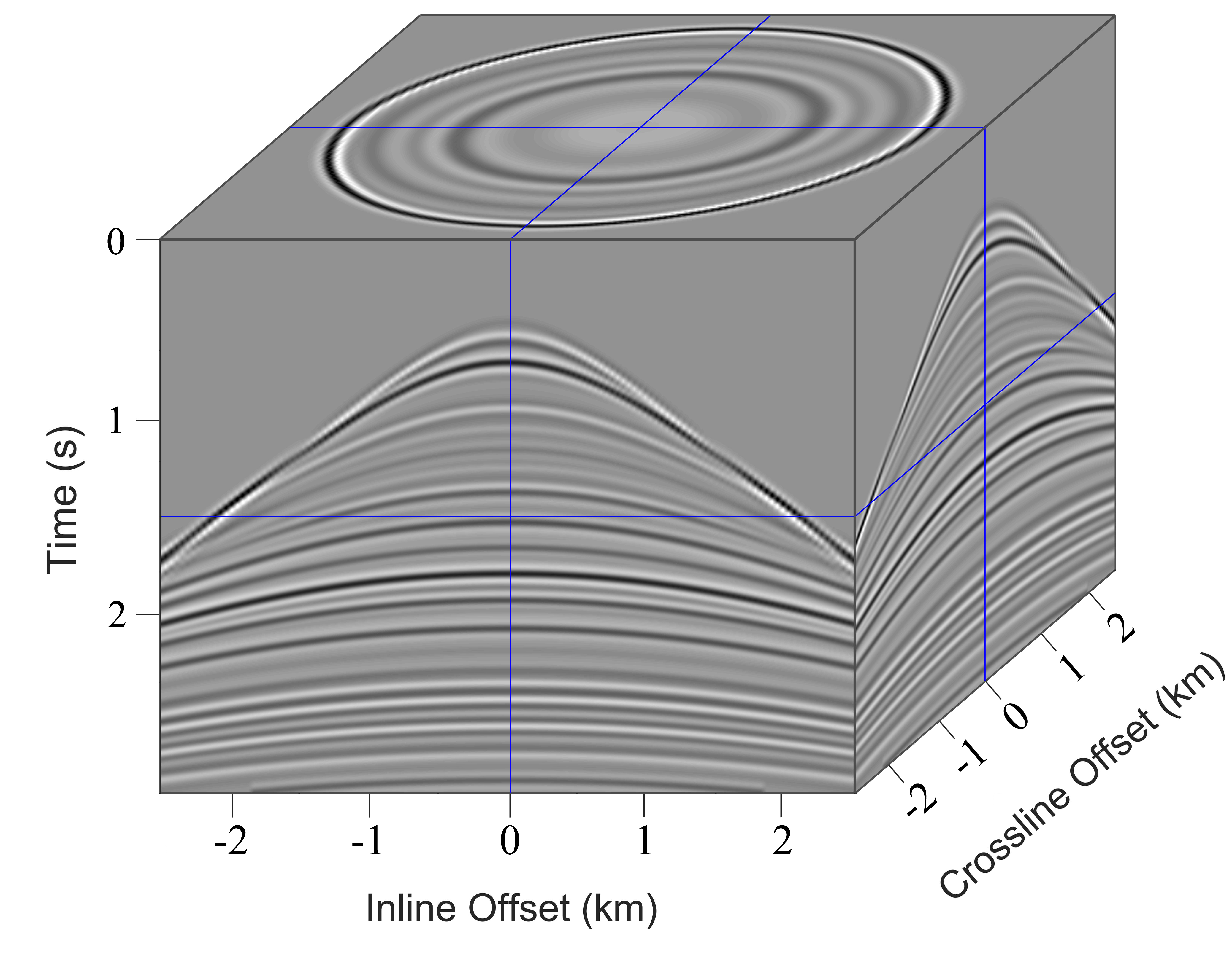}
\caption{Synthetic 3D CMP gather with azimutal velocity anisotropy.}
\label{Fig_syn_3d}
\end{figure}

\begin{figure}
\centering
\includegraphics[width=13cm,height=12cm]{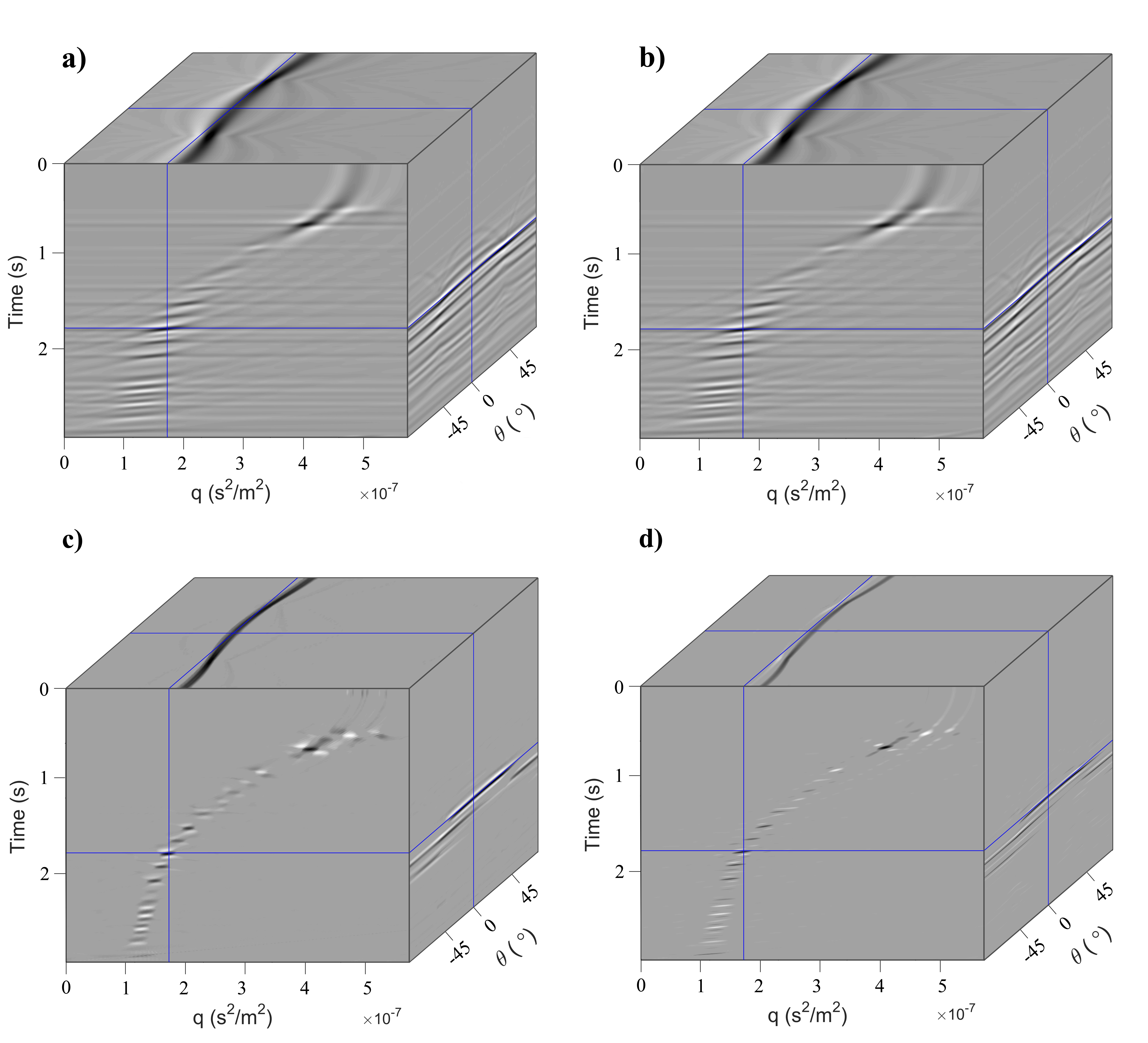}
\caption{The results of azimuthally anisotropic 3D velocity analysis of the CMP in Fig. \ref{Fig_syn_3d} via (a) direct summation and (b) proposed proposed algorithm. (c) The results of sparse velocity stack. (d) The results of sparse deconvolutive velocity stack. }
\label{Fig_RTs}
\end{figure}

\begin{figure}
\centering
\includegraphics[width=7cm,height=6cm]{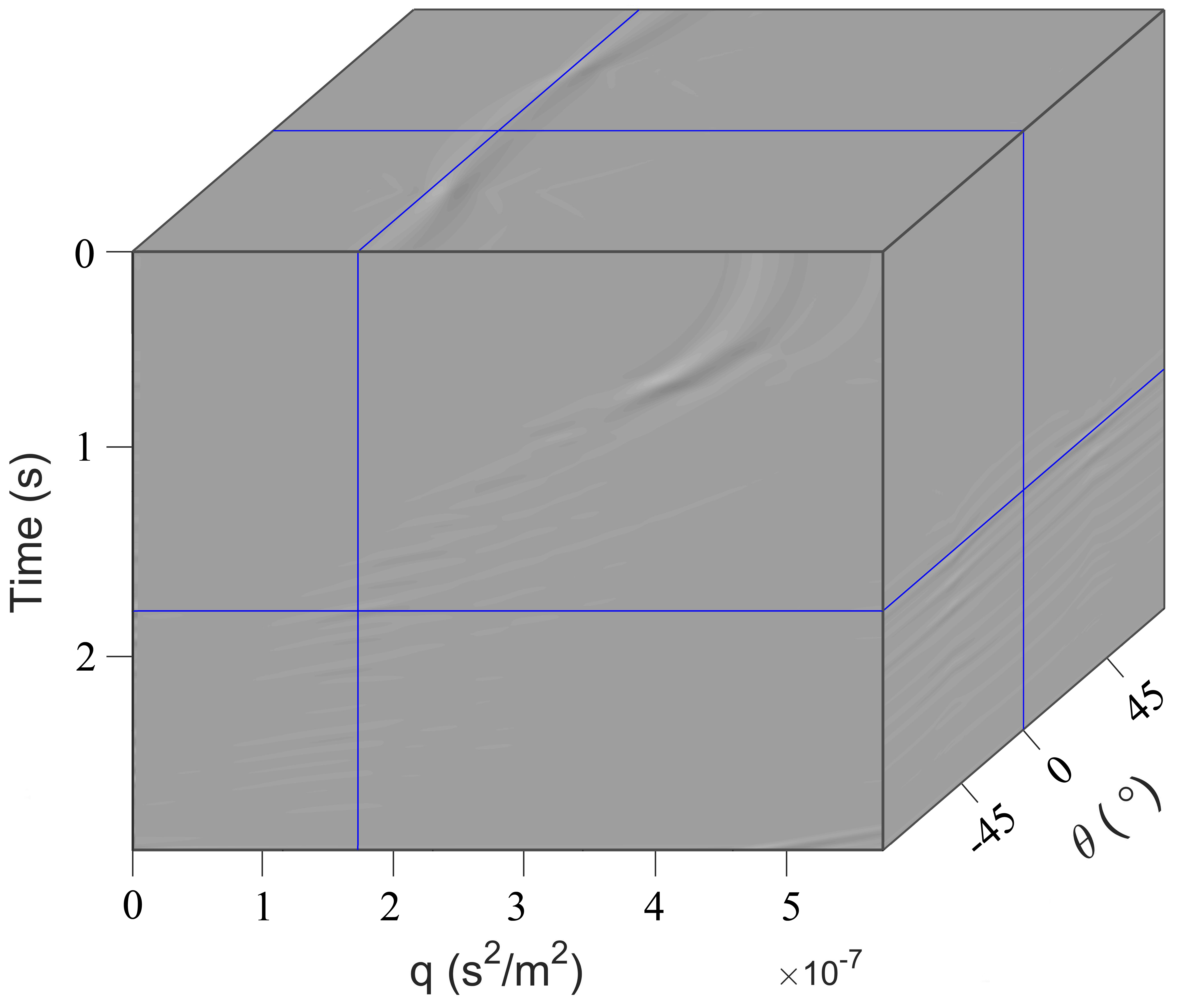}
\caption{The difference between AAHRTs via direct summation (figure \ref{Fig_RTs}-a) and proposed algorithm (figure \ref{Fig_RTs}-b). The same scale as in figure \ref{Fig_RTs}-a is used to plot the cube.}
\label{Res_direct_proposed}
\end{figure}

We also transformed the synthetic data in figure \ref{Fig_syn_3d} by using the iterative sparse solver given in eq. \eqref{ISTA}.
We applied the algorithm both in its traditional form and the deconvolutiove form. In the latter case, we used the true wavelet (a 10 Hz ricker wavelet) to construct the basis functions and hence the operator in eq. \eqref{DecRT}.
Figures  \ref{Fig_RTs}c and \ref{Fig_RTs}d respectively show the result obtained by sparse velocity stack in the traditional (Fig. \ref{Fig_RTs}c) and deconvolutive (Fig. \ref{Fig_RTs}d) forms.  resolution and high resolution deconvolutive AAHRTs are shown in figure \ref{Fig_RTs}-b to d, respectively. Also, in figure \ref{Fig_RTs}-a, low resolution AAHRT via direct summation with linear interpolation in time and azimuth is shown.
Comparing \ref{Fig_RTs}a-b and \ref{Fig_RTs}c-d clearly shows higher resolution of the coefficients due to the sparsity constraint. Furthermore, it is seen that the modified basis functions in the deconvolutive transform improves the sparsity of the coefficients. (Fig. \ref{Fig_RTs}d versus \ref{Fig_RTs}c).

\begin{figure}
\centering
\includegraphics[width=13cm,height=10cm]{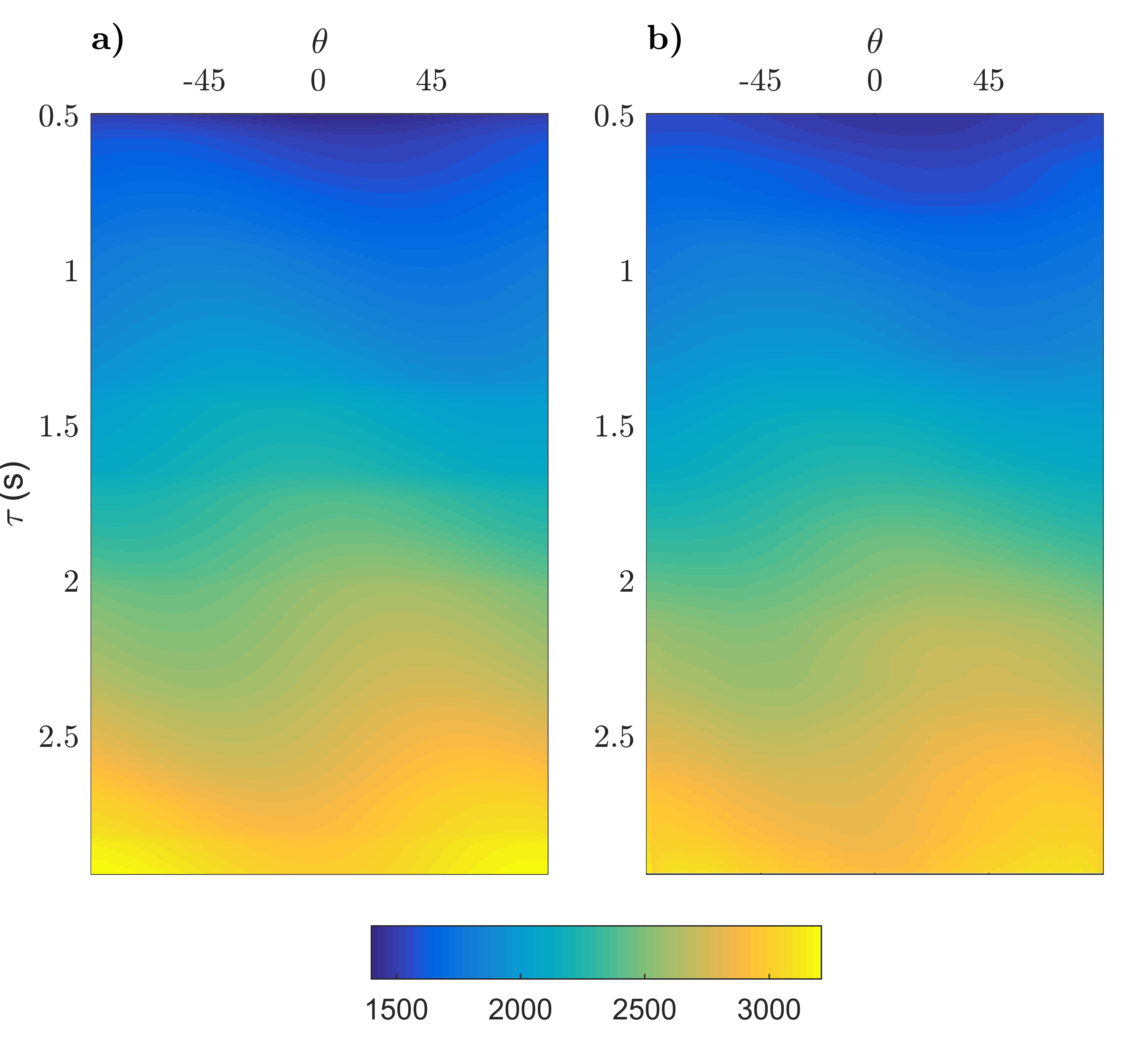}
\caption{a) The velocity vs azimuth used to construct the synthetic data in figure \ref{Fig_syn_3d}. b) The estimated velocity vs azimuth via proposed method. }
\label{Fig_vel_compare}
\end{figure}

\begin{figure}
\centering
\includegraphics[width=8cm,height=11cm]{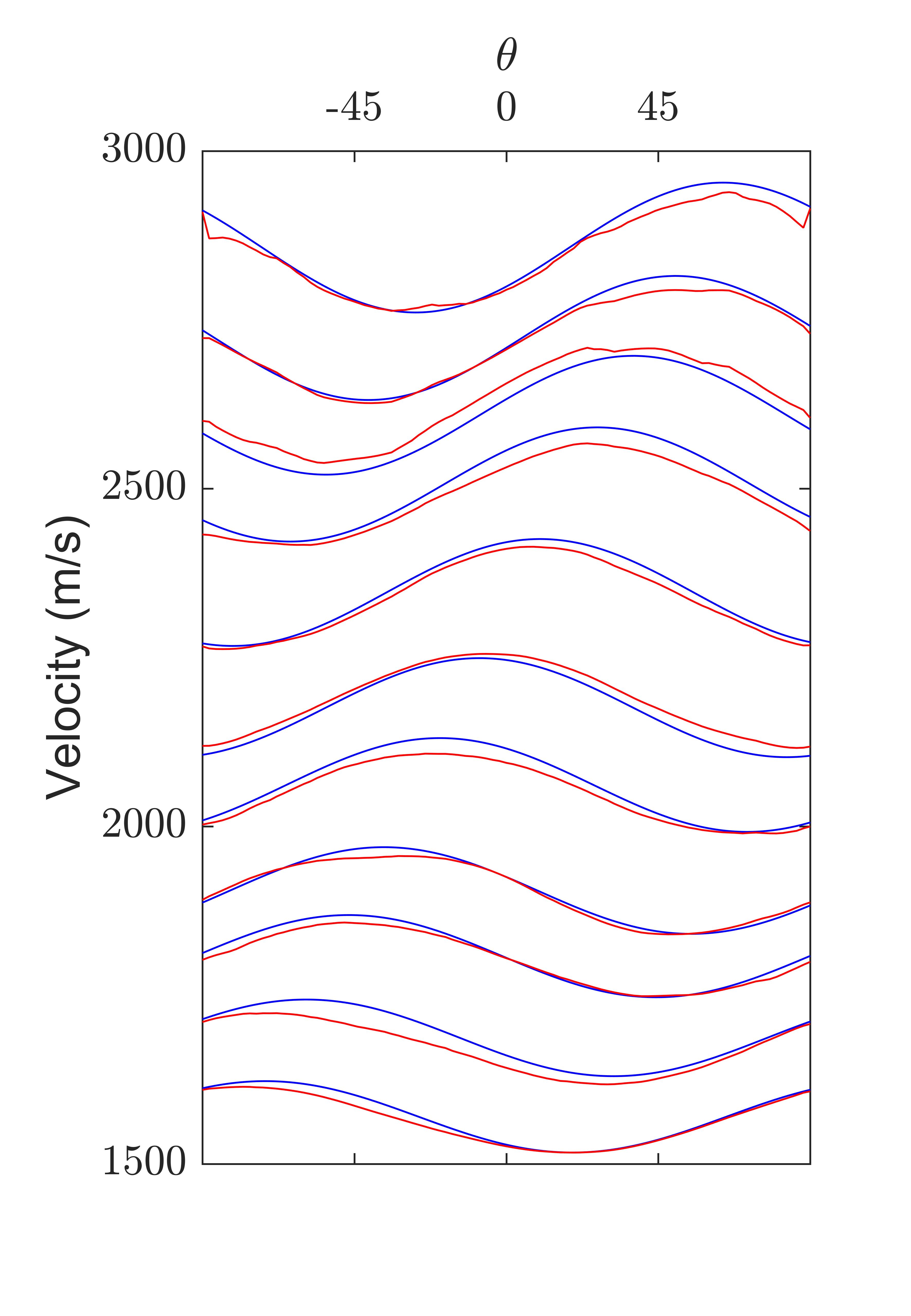}
\caption{(Blue) true and (red) estimated stacking velocities versus azimuth. From bottom to top, the curves correspond to $\tau=0.6,~0.8,~1,~1.2,~1.4,~1.6,~1.8,~2,~2.2,~2.4$ and $2.6$ s.}
\label{vel_est_tau}
\end{figure}

We estimated the stacking velocities (as a function of time and azimuth) from the deconvolutive Radon cube in Fig. \ref{Fig_RTs}d by using eq. \ref{vel_est} and the result is shown in Fig. \ref{Fig_vel_compare}b. For a better comparison, the estimated and true velocities are plotted at different depths in Fig. \ref{vel_est_tau}.
It is seen that the stacking velocity are estimated with good accuracy.

\begin{figure}
\centering
\includegraphics[width=14cm,height=5cm,trim={0 0 20cm 0},clip]{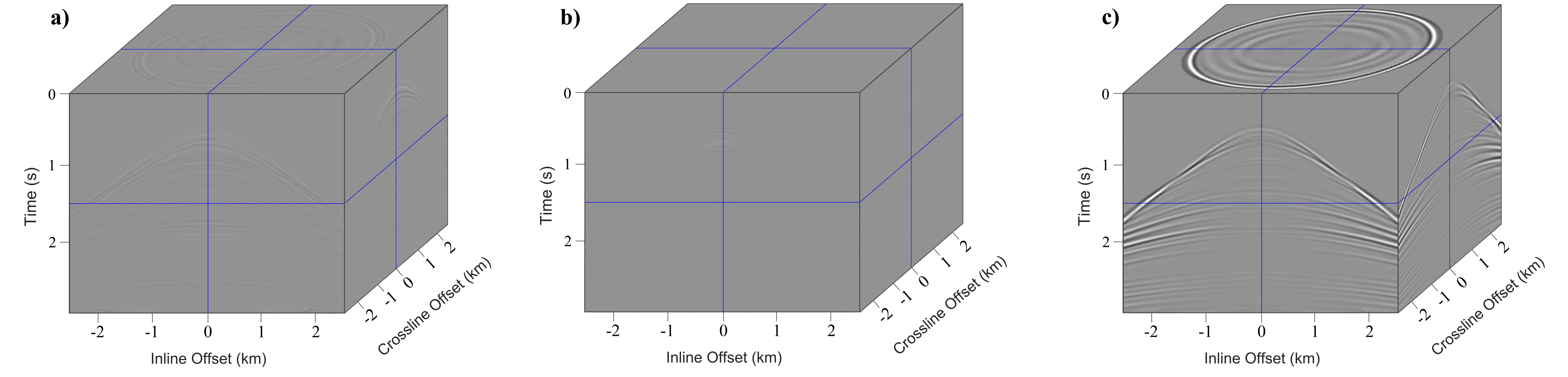}
\caption{Difference between the CMP data in Fig. \ref{Fig_syn_3d} and the reconstructed data by (a) traditional AAHRT and (b) deconvolutive AAHRT.}
\label{Res_recons}
\end{figure}

We also compared different methods in the data domain by calculating the residuals (the difference between observed and calculated data).
Figures \ref{Res_recons}a and \ref{Res_recons}b show the residual data corresponding to the traditional and deconvolutive AAHRTs. As can be seen, the data is well fitted by both methods. A comparison between the Radon cubes \ref{Fig_RTs}c and \ref{Fig_RTs}d and the corresponding data residuals \ref{Res_recons}a and \ref{Res_recons}b clearly confirm that the deconvolutive transform fits the data with significantly less basis functions. This makes this transform suitable for anisotropic data reconstruction.



\begin{table}
\caption{Computational times to apply azimuthally anisotropic 3D velocity scan in seconds.}
\centering
\begin{tabular}{c c c c  }
\hline\hline
$(N_t,N_x,N_y,N_p,N_\theta)$ & direct sum  & proposed method  &  speedup \\ [0.5ex]
\hline
$ (512,128,128,128,90) $    & 218     & 0.87 & 250  \\
$ (512,128,128,256,90) $    & 421     & 1.14 & 369  \\
$ (512,256,256,128,90) $    & 915     & 3.80 & 240  \\
$ (512,256,256,256,90) $    & 1685    & 3.98 & 423  \\
$ (1024,128,128,128,90) $   & 497     & 1.97 & 252  \\
$ (1024,128,128,256,90)$    & 875     & 2.78 & 314  \\
$ (1024,256,256,128,90) $   & 1806    & 8.57 & 210  \\
$ (1024,256,256,256,90)$    & 3416    & 8.82 & 387  \\
\hline
\end{tabular}
\label{table:table1r}
\end{table}


\section{Conclusions}
The Generalized Fourier Slice Theorem (GFST) establishes an analytical relationship between the Fourier representation of a signal and the Fourier representation of its Radon transform. This relationship enables efficient computation of Radon transforms using interpolation techniques combined with the Fast Fourier Transform (FFT). The results presented in this paper demonstrate that the high-resolution, fully anisotropic 3D Radon transform can be computed using GFST up to 400 times faster than with conventional direct methods.
\bibliographystyle{gji}
\newcommand{\SortNoop}[1]{}

\end{document}